\documentclass{article}
\usepackage[T1]{fontenc} 
\usepackage[utf8]{inputenc} 
\usepackage{ismir,amsmath,cite,url}
\usepackage{graphicx}
\usepackage{color}
\usepackage{subcaption}
\usepackage{caption}
\usepackage{url}
\usepackage{enumitem}
\usepackage{tabu}
\usepackage[bookmarks=false]{hyperref}
\hypersetup{colorlinks,allcolors=black}
\usepackage{textcomp}
\usepackage{lineno}

\title{Learning long-term music representations via hierarchical contextual constraints}

\multauthor
{Shiqi Wei$^{1,2}$  \hspace{1cm}  Gus Xia$^2$}
 { $^1$ School of Data Science, Fudan University\\
$^2$ Music X Lab, Computer Science Department, New York University Shanghai\\
{\tt\small sqwei19@fudan.edu.cn, gxia@nyu.edu}
}

\def\authorname{Shiqi Wei, Gus Xia}

\begin{document}

\maketitle
\begin{abstract}
Learning symbolic music representations, especially disentangled representations with probabilistic interpretations, has been shown to benefit both music understanding and generation. However, most models are only applicable to short-term music, while learning long-term music representations remains a challenging task. We have seen several studies attempting to learn hierarchical representations directly in an end-to-end manner, but these models have not been able to achieve the desired results and the training process is not stable. In this paper, we propose a novel approach to learn long-term symbolic music representations through contextual constraints. First, we use contrastive learning to pre-train a long-term representation by constraining its difference from the short-term representation (extracted by an off-the-shelf model). Then, we fine-tune the long-term representation by a hierarchical prediction model such that a good long-term representation (e.g., an 8-bar representation) can reconstruct the corresponding short-term ones (e.g., the 2-bar representations within the 8-bar range). Experiments show that our method stabilizes the training and the fine-tuning steps. In addition, the designed contextual constraints benefit both reconstruction and disentanglement, significantly outperforming the baselines. 
\end{abstract}

\section{Introduction}\label{sec:introduction}

Deep music representation learning have been proven to be a powerful tool for high-quality symbolic music generation \cite{DBLP:journals/corr/abs-1803-05428}. The learned representations can be directly fed into downstream predictive models such as LSTMs\cite{lstm} and Transformers\cite{attention} to achieve more coherent results than note-based or event-based generation \cite{DBLP:journals/corr/abs-2008-01291,DBLP:journals/corr/abs-2008-07122,Inpainting} 
Furthermore, when a representation learning model has a probability interpretation, the representation can then  be easily interpolated or resampled to create new music pieces. Recently, several studies further disentangle music representations into interpretable factors (such as pitch, rhythm, chord and texture) to achieve a more controllable and interactive music generation \cite{DBLP:journals/corr/abs-1906-03626,DBLP:journals/corr/abs-2008-07122,poptransformer}. For example, we can keep the pitch factor of a melody while resampling its rhythm factor to achieve theme variation. We can also interpolate the pitch factor for a smooth music morphing \cite{DBLP:journals/corr/abs-1906-03626}. 

Despite the above mentioned progress \cite{midivae,DBLP:conf/ismir/Akama19,DBLP:conf/ismir/LuoAH19}, most existing work applies only to short music segments with a length of several beats, while learning long-term representations remains a challenging task. In particular, studies have shown that even for monophonic melodies,  "flat" model designs (e.g., using long-range sequential encoders) have difficulty remembering a complete music phrase at once. Some other studies have attempted to solve this problem by building another layer of hierarchy on top of short-range flat models, learning short-term and long-term representations simultaneously in an end-to-end manner \cite{DBLP:journals/corr/abs-1803-05428,9054554}. However, as the model expressivity increases with the number of layers, models also become much more difficult to train.

We argue that the main problem with current methods is the lack of proper inductive bias, and in this paper we propose a new method for learning long-term, phrase-level symbolic music representations through contextual constraints. The method consists of two stages pre-training and fine-tuning, with two steps in each stage. In the pre-training stage, we first adopt EC$^2$-VAE \cite{DBLP:journals/corr/abs-1906-03626} to learn bar-level, disentangled latent pitch and rhythm representations. Then, we apply the same model to learn phrase-level representation but with contrastive losses to constrain the difference between phrase-level and bar-level representations. It is indeed difficult to learn phrase-level representations directly using  bar-level models, but the additional contrastive constraint can serve as a useful inductive bias to help find a reasonable solution that can subsequently be improved by fine-tuning. During the fine-tuning stage, we replace the pre-trained decoder with a hierarchical prediction model that forces the phrase-level representation to reconstruct the bar-level ones. This is achieved by first tuning only the new hierarchical decoder (while fixing the pre-trained encoder) and then tuning the whole network. During these two steps, structured contrastive loss is applied to stabilize the learning process.

Experiments show that the proposed method significantly outperforms the baselines and successfully learns disentangled pitch and rhythm representations for 8-bar long phrases (32 beats in 4/4 meter) without increasing the latent dimensionality. To our knowledge, this is also the first generative model that achieves phrase-level composition style transfer, latent factor interpolation, and theme variation. In sum, our contributions are as follows:
\begin{itemize}[parsep=0pt, itemsep=0pt,leftmargin=*]
    \item We demonstrate the importance of structured contextual constraints in learning long-term disentangled representations. Our approach only requires reasonable amount of data to train and could learn compact latent representation.
    \item We show that the proposed Structured InfoNCE loss effectively expresses the contextual constraints, stabilizes the training of long-range models and helps the model converge faster.
    \item Our model achieves phrase-level music style transfer, latent factor interpolation, and theme variation.
\end{itemize}

\section{Related Work}
We review two realms of research related to our work on long-term music-representation learning: contrastive learning, which is the main method to stabilize the training process, and hierarchical music modeling, which is related to our fine-tuning model.
\subsection{Contrastive Learning}
Contrastive learning (CL) is an efficient method in self-supervised learning\cite{DBLP:conf/icml/ChenK0H20,DBLP:conf/cvpr/He0WXG20,DBLP:journals/corr/abs-2003-04297}, serving as regularization to latent representations. For example, NCE-based contrastive losses\cite{ncebased,WORDNCE} have been widely used and achieved good results in natural language processing. Contrastive predictive coding (CPC) \cite{DBLP:journals/corr/abs-1807-03748} and Deep Infomax (DIM) \cite{DBLP:conf/iclr/HjelmFLGBTB19} explore the relation between minimizing a contrastive learning loss and maximizing a lower bound of the mutual information. In DIM, global feature is connected with local feature to learn more abstract and informative representations. 

\subsection{Hierarchical Music Representation Learning}

The hierarchical nature of music has been studied for a long time\cite{1996A,1983Introduction,2006Implementing,Schenkerian}. Recently, we see some efforts on learning long-term music representations using hierarchical modeling\cite{MusicFaderNets,VQmusic,9054554}. The basic idea is that since a flat model design can only effectively learn shot-term representations, we can stack more layers on top of the short-term representations module for long-term representations. Existing works include MusicVAE \cite{DBLP:journals/corr/abs-1803-05428}, Music Transformer VAE \cite{9054554}, Jukebox \cite{dhariwal2020jukebox}, etc. However, experiments show that unless we have a huge amount of data, the model is in general very difficult to train. In this study, we provide a two-stage algorithm with contrastive loss as a better learning strategy. Also, no model so far has achieved disentanglement for long-term representation as done in this study.

\section{Methodology}
In this section, we introduce our algorithm in detail. Conceptually, it consists of two stages, each with two steps.The first stage is \textit{pre-training}:
\begin{itemize}[parsep=0pt, itemsep=0pt]
    \item In step 1, we simply adopt EC$^{2}$-VAE \cite{DBLP:journals/corr/abs-1906-03626}, an existing music representation disentanglement model, to extract short-term pitch and rhythm representations.
    \item In step 2, we build Long-EC$^{2}$-VAE, a long-term version of the model and train it with an extra contextual constraint using the proposed \textit{Structured} InfoNCE loss. Intuitively, this loss prevents the learned long-term representations from deviating too far from corresponding well-trained short-term representations.
\end{itemize}
The second stage is \textit{fine-tuning}, in which we build a hierarchical representation-learning model by combining the encoder of Long-EC$^{2}$-VAE with a hierarchical decoder. We name this model after Hierarchical-EC$^{2}$-VAE. 
\begin{itemize}[parsep=0pt, itemsep=0pt]
    \item In step 1, we only train the hierarchical decoder to ensure the predictive power of the long-term representation.
    \item In step 2, we train the whole hierarchical network for a better long-term pitch-rhythm disentanglement. 
\end{itemize}

\subsection{Pre-training by Contrastive Learning}
\begin{figure}[htbp]
 \centerline{
 \includegraphics[width=0.9\linewidth]{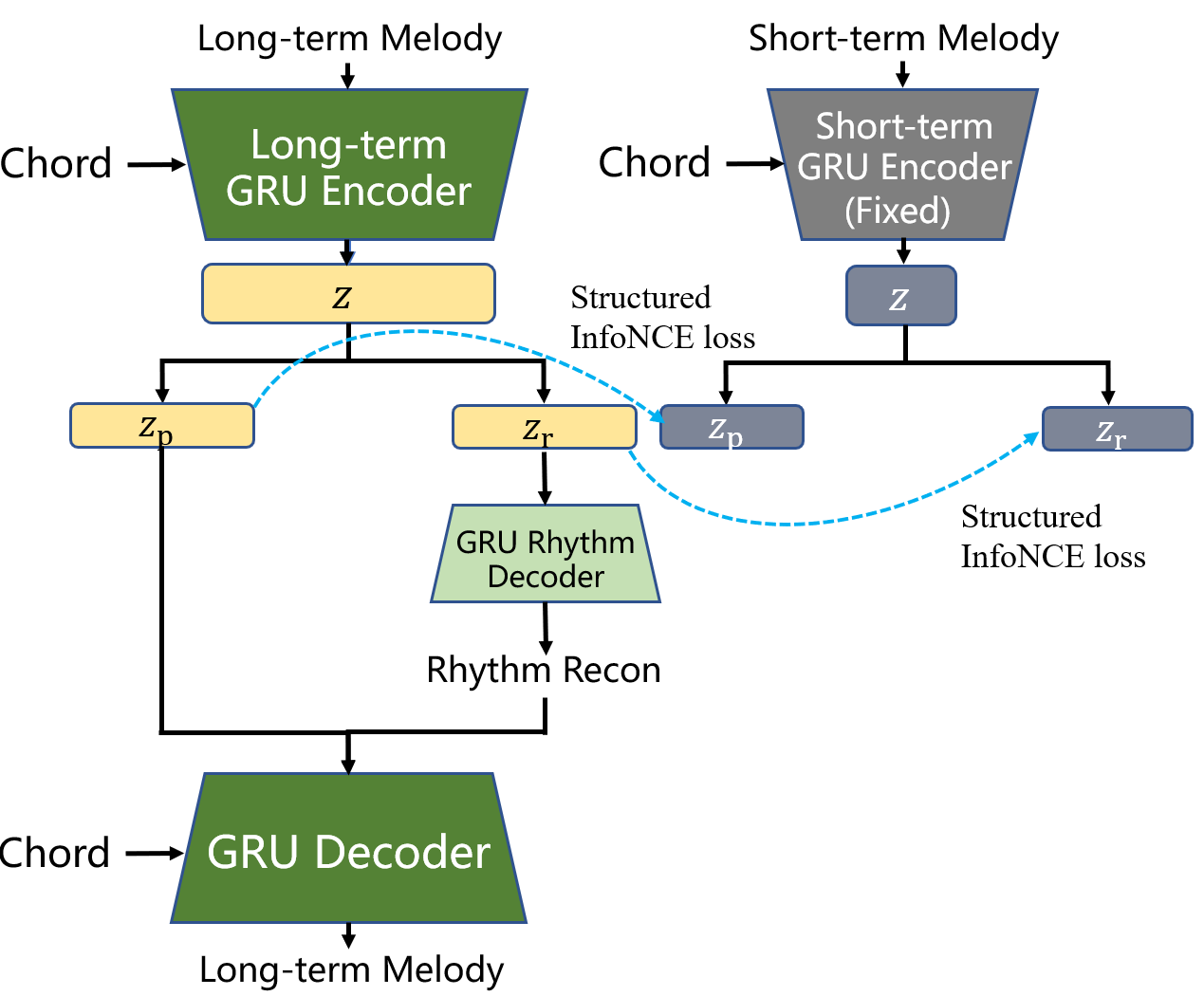}}
 \caption{The model architecture of Long-EC$^{2}$-VAE in the pre-training stage, where the right-hand-side is short-term model with parameter fixed and the left-hand-side is the long-term model. The dotted lines denote contrastive losses, whose weighting matrices are joined optimized with the parameters on the left-hand-side networks.}
 \label{fig:step1}
\end{figure}

The model of the pre-training stage, Long-EC$^{2}$-VAE, is shown in Figure \ref{fig:step1}. It is built upon an off-the-shelf music representation model, EC$^{2}$-VAE  \cite{DBLP:journals/corr/abs-1906-03626}, which can effectively disentangle pitch and rhythm factors for short music segments by cutting the latent representation into two parts and pairing one part with a local rhythm decoder. 
\begin{figure*}[t]
    \centering
        \includegraphics[width=\linewidth]{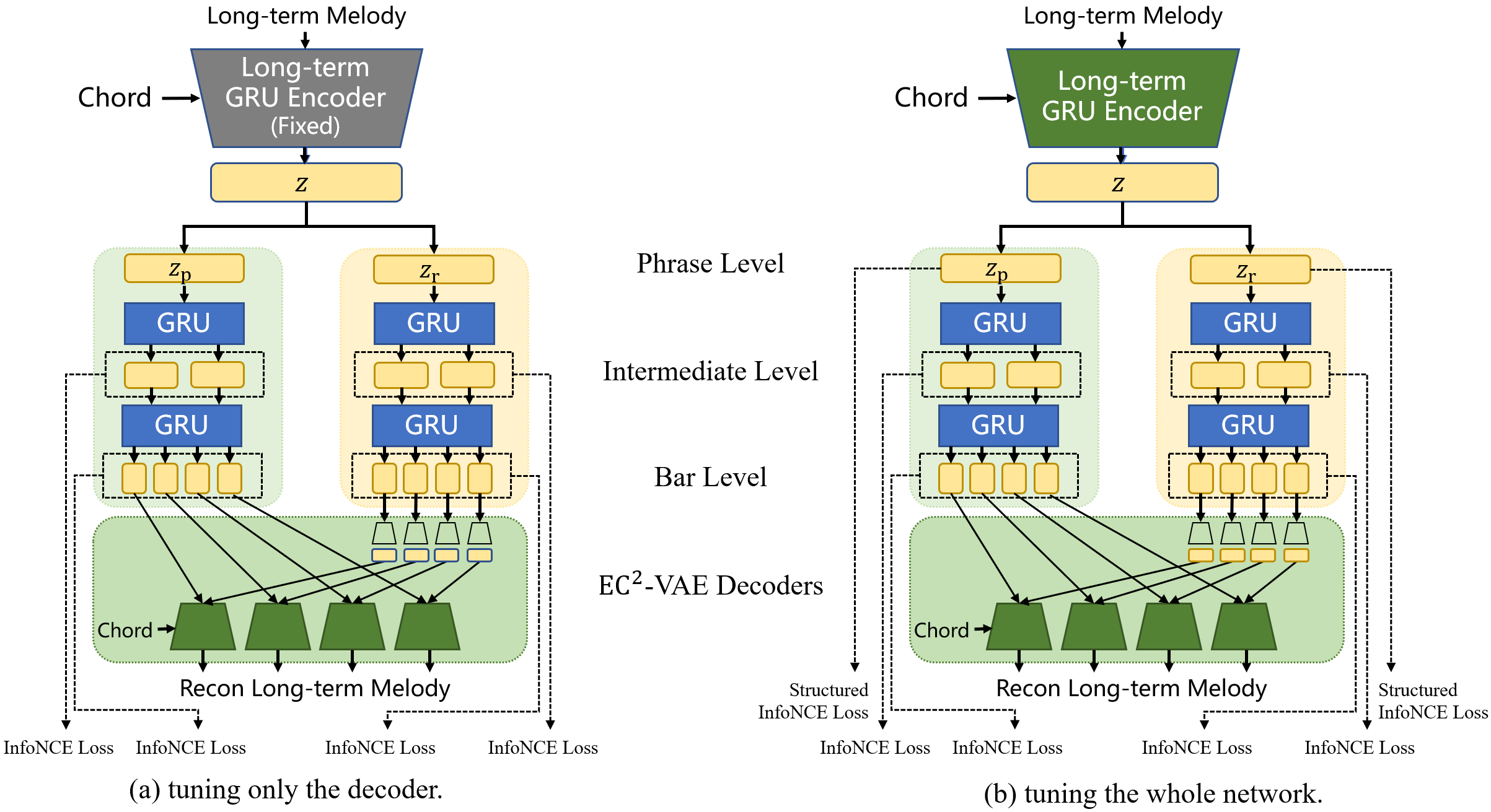}
    \caption{The model architecture of  Hierarchical-EC$^2$-VAE in the fine-tuning stage. The training follows two steps.}
    \label{fig:compare_bar_kld}
\end{figure*}
In Figure~\ref{fig:step1}, the right-hand-side part is a literal copy of the EC$^{2}$-VAE encoder (with parameters fixed) to extract short-term representations, while the left-hand-side part is a simple adaptation of EC$^{2}$-VAE for long-term music by lengthening its temporal receptive field. Note that the left part alone is not able to learn long-term representations, and our goal is to assist it using contrastive learning. Formally, the loss function of Long-EC$^{2}$-VAE is:
\begin{equation}
\label{eq:total_loss}
\mathcal{L} = \mathcal{L}_{\rm Long\text{-}EC^{2}\text{-VAE}}+\mathcal{L}_{\rm Structured\, InfoNCE},
\end{equation}
where $\mathcal{L}_{\rm Long\text{-}EC^{2}\text{-VAE}}$ is the same as in the original EC$^{2}$-VAE model (which contains the KL loss, the rhythm loss and the overall reconstruction loss). The Structured InfoNCE loss expresses the contextual constraint. It is developed from InfoNCE \cite{DBLP:journals/corr/abs-1807-03748} loss, and it is \textit{structured} since the compared representation pairs are extracted from music segments of different length, one is long term and the other is short term. Formally:

{\footnotesize
\begin{equation}
\begin{aligned}
&\mathcal{L}_{\text{Structured InfoNCE}} =\\
&-\ln \frac{\exp{\left(z^{T}_{\text{L},f} W \hat{z}^{+}_{\text{ S},f}/\tau\right)}}{\exp{\left(z^{T}_{\text{L},f} W  \hat{z}^{+}_{\text{ S},f}/\tau\right)}+\sum_{i=1}^{K}\exp{\left(z^{T}_{\text{L},f} W  \hat{z}^{-}_{\text{S},f}/\tau\right)}},\label{str.Info}
\end{aligned}
\end{equation}
}
where $z_{\text{L}, f}$ and $W$ are the normalized long-term representations and weighing matrix we need to learn. $f =\{\rm{p}, \rm{r}\}$ indicates whether it is the pitch or rhythm factor. Likewise, we use $\hat z_{\text{S},f}$ to denote the short-term representations extracted by right-hand-side model.
$K$ and $\tau$ are hyper-parameters. $K$ is the amount negative samples and $\tau$ is the temperature parameter.  

In specific, the short-term melodies are half as long as long-term ones. The positive samples $\hat{z}^{+}_{\text{ S}, f}$ are in the cases that the corresponding short-term melody is a part of the long-term melody and $f$ takes the same value as in ${z}_{\text{L}, f}$, while the negative samples are not in this case. Also, the  long-term and short-term representations share the \textit{same} dimensionality. 

\subsection{Fine-tuning with Hierarchical Generation}
Figure \ref{fig:compare_bar_kld} shows the architecture of the fine-tuning model, Hierarchical-EC$^{2}$-VAE, where the two subfigures illustrate the two training steps. Here, the encoder design is the same as in the Long-EC$^{2}$-VAE model, while the decoder is a hierarchical predictive model with three layers. The first two layers are new designed and the last layer is an aggregation of several EC$^{2}$-VAE decoders sharing the same parameters. Given the disentangled long-term (phrase-level) representations, it first decodes intermediate-level representations, then decodes bar-level representations, and finally reconstructs concrete rhythm and music tokens. 

Compared to the phrase-level representation, the temporal receptive fields of the intermediate-level representations all shrink to a half, but at the same time their number doubles in order to cover the same range of music. The same relationship holds between intermediate and bar-level representations. In particular, a phrase means 8 bar (in 4/4 meter, 32 beats) in our design, so that the intermediate-level and bar-level mean 4-bar and 2-bar melody segments (a length which the original EC$^2$-VAE model can handle), respectively. All levels of latent representations share the same dimensionality.

In the first step of training (Figure \ref{fig:compare_bar_kld}(a)), the encoder is a literal copy from the Long-EC$^{2}$-VAE model and we only train the hierarchical decoder. Formally, the loss function is:
\begin{equation}
\begin{aligned}
    \mathcal{L}_{\rm step 1} = \mathcal{L}_{\rm Hierarchical-EC^{2}\text{-VAE Decoder}} +\mathcal{L}_{\rm InfoNCE},\label{step1eq}
\end{aligned}
\end{equation}
where the first term refers to the reconstruction losses adopted from the EC$^2$-VAE model, and the second term is defined as:

\begin{footnotesize}
\begin{equation}
\begin{aligned}
&\mathcal{L}_{\rm InfoNCE} =\\
&-\ln \frac{\exp{\left(z^{T}_{l,f}W \hat{z}^{+}_{l,f}/\tau\right)}}{\exp{\left(z_{l,f}^{T}W \hat{z}^{+}_{l,f}/\tau\right)}+\sum_{i=1}^{K}\exp{\left(z^{T}_{l,f} W\hat{z}^{-}_{l,f}/\tau\right)}},\label{info}
\end{aligned}
\end{equation}
\end{footnotesize}
where  $z_{l, f}$ are the normalized hierarchical representations we need to learn with  $l =\{\rm{intermediate}, \rm{bar}\}$ indicating the level of representation and other notations follow the same meaning as in  Eq.(\ref{str.Info}). Here, both positive $\hat{z}^{+}_{l,f}$ and negative $\hat{z}^{-}_{l,f}$ samples are normalized representations computed from a pre-trained EC$^{2}$-VAE, in which the positive samples are in the cases that the $\hat{z}^{+}_{l,f}$ and $z_{l, f}$ are computed based on the same music segment and have the same value of $l$ and $f$, while  $\hat{z}^{-}_{l,f}$ are not in this case. 

After the first step achieves a reasonable accuracy, we proceed to step 2 (Figure~\ref{fig:compare_bar_kld}(b)), unfreezing the encoder and training the whole hierarchical representation-learning model with:
\begin{equation}
\begin{aligned}
    \mathcal{L}_{\rm step 2} = \mathcal{L}_{\rm step 1}&+\mathcal{L}_{\rm Structured\, InfoNCE}\\
    &+\beta\mathcal{L}_{\rm KL\, phrase},\label{step2eq}
\end{aligned}
\end{equation}
where the first two terms are defined in Eq.~(\ref{step1eq}) and Eq.~(\ref{str.Info}) respectively. $L_{\rm KL\, phrase}$ is KL divergence to only regularize the phrase-level representations by a normal distribution. The value $\beta$ controls the degree of KL divergence penalty. 
\section{Experiments}
\subsection{Dataset and data format}
We train our model on Nottingham Database\cite{Nottingham} and POP909 database \cite{DBLP:journals/corr/abs-2008-07142}. 
Our dataset contains 2154  melodies (at song level) in total.
We randomly split these pieces into 2 subsets: $90\%$ pieces for  training and for $10\%$ pieces for test.
The data format is designed as the same as in \cite{DBLP:journals/corr/abs-1906-03626} in which  4 bar or 8 bar melodies are formalized as  sequences of 130-dimensional one-hot embedding vectors and 16-beat and 32-beat rhythm pattern is represented by a sequence of  3-dimensional one-hot embedding vectors. Each vector in the melody sequence denotes a $\frac{1}{4}$-beat unit. The first 128 dimensions of this vector denote 128 MIDI-format pitches from $0$ to $127$, the $129^{\rm th}$ dimension is the holding state for longer note duration, and the last dimension is kept for rest. The three dimensions of rhythm pattern vectors represent the onset of any pitch, a holding state, and rest, respectively.

\subsection{Implementation Details}
All of our models are trained using Adam optimizer \cite{DBLP:journals/corr/KingmaB14} with a scheduled learning rate from 1e-3 to 1e-5. The batch size is 128 in the pre-training stage and is 64 in the fine-tuning stage. We do normalization on representations in Eq.(\ref{str.Info}) and (\ref{info}) to make the training process more stable. The representations fed into decoders are original representation without normalization. 

\subsubsection{Pre-training}
In the pre-training stage, we simply adopt the structure of  EC$^{2}$-VAE~\cite{DBLP:journals/corr/abs-1906-03626} to model 4 bar and 8 bar EC$^{2}$-VAE.  Each  model comprises an encoder with a bi-directional GRU layer, a rhythm decoder with a GRU layer, and a global decoder with a GRU layer. We set the hidden dimension of the GRU in the encoder and decoders to 2048. The latent dimension is 128 for disentangled pitch representations and 128 for disentangled rhythm representations for each range model. 
For $\mathcal{L}_{\text{Structured InfoNCE}}$ depicted in Eq.~(\ref{str.Info}), we set $ K $ to 512 and $\tau$ to 1.  The positive samples for Eq.~(\ref{str.Info}) and Eq.~(\ref{info}) are the representations of 1-4th, 3-6th and 5-8th bar from well-trained 4 bar EC$^2$-VAE. Actually, even when training the 4-bar EC$^2$-VAE (right-hand side of Figure 1), we use a similar constrastive loss as in Eq.~(\ref{str.Info}) where the positive samples are representations of 1-2th, 2-3th, 3-4th bar from well-trained 2 bar (original) EC$^2$-VAE~\cite{DBLP:journals/corr/abs-1906-03626} .

\subsubsection{Fine-tuning}
 Hierarchical-EC$^{2}$-VAE model consists of a long-term (8 bar) EC$^2$-VAE encoder, 4 GRU layers, and an aggregation of 2 bar EC$^{2}$-VAE decoders. We first train the hierarchical model with fixed 8 bar EC$^2$-VAE encoder from pre-trained stage for around 25 epochs. Then we train the whole model without fixing  parameters.  We set the hidden dimension of  4 GRU layers to 1024. We set $ K $ to 256 and  $\tau$ to 1 for both Structured InfoNCE loss and InfoNCE loss and set $\beta$ to 0.1 in Eq.~(\ref{step2eq}).

\subsection{Objective Evaluation}
We objectively evaluate the model in terms of reconstruction accuracy, training stability, and disentanglement.
\subsubsection{Reconstruction Accuracy}
Table~\ref{table:acc} shows that the reconstruction accuracy of the proposed models (2nd an 3rd rows) significantly outperform the baseline, a vanilla EC$^2$-VAE applied to 8-bar melody (first row). The last two rows show the results of two ablation settings of Hierarchical-EC$^{2}$-VAE: one without the contrastive loss and the other without first fixing the parameters of encoder and directly train the model end-to-end. We see that the proposed Structured InfoNCE or InfoNCE losses play a vital role for an accurate reconstruction and the two-step training strategy improves the result marginally.
{\footnotesize
\begin{table}[htp]
    \centering
    \setlength{\tabcolsep}{1.0mm}{
    \begin{tabular}{c|c|c} \hline 
      Method & Recon.Acc& Rhythm Recon.Acc \\ 
      \hline
      Baseline   & 0.772 &0.847\\ 
      Long-EC$^{2}$-VAE & \textbf{0.992} &\textbf{0.995}\\ 
      H-EC$^{2}$-VAE(ours)  & \textbf{0.995}&\textbf{0.995} \\ 
      H-EC$^{2}$-VAE(w/o CL) & 0.584 & 0.599 \\ 
      H-EC$^{2}$-VAE(w/o fixed) & 0.991 & 0.989 \\ 
      \hline 
    \end{tabular} }
    \caption{A comparison on reconstruction accuracy of different models.}
    \label{table:acc}
\end{table}}
\subsubsection{Training stability}
\begin{figure}[htp]
        \centering
        \hfill
         \includegraphics[width=\linewidth,height=3.5cm,clip, trim=0cm 1.5cm 0cm 2cm]{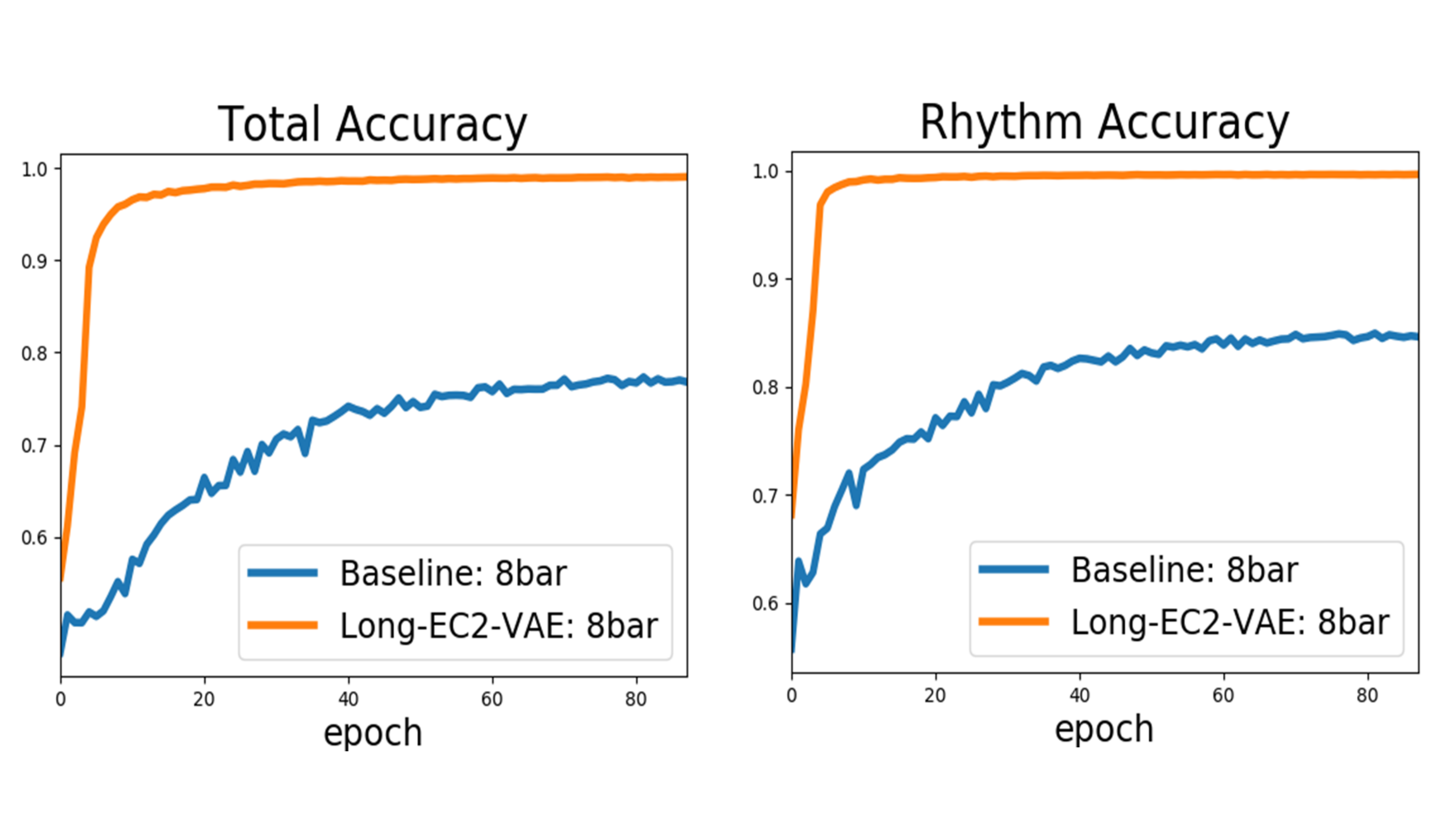}
    \caption{Experimental results of  overall reconstruction and rhythm accuracy on the test set.}
    \label{fig:acc}
   \vspace{-1em}
\end{figure}
Comparing the accuracy curves of the proposed Long-EC$^{2}$-VAE with the baseline as illustrated in Figure~\ref{fig:acc}, we find that the proposed Long-EC$^{2}$-VAE converges more quickly during training. This indicates that the proposed training strategy leads to a better initialization and makes the performance of the model fluctuate less during training.

\subsubsection{Disentanglement Evaluation}
We evaluate the disentanglement performance of models using a disentanglement evaluation method adopted in~\cite{DBLP:journals/corr/abs-1906-03626} and~\cite{DBLP:conf/icml/KimM18}. The method randomly transposes all the notes of the input data by $i (i \in [1, 12])$ semitones while keeping the rhythm and underlying chord unchanged and then measures the variation of disentangled representations.We denote $\Sigma|\Delta z_{\text{p}}|$ and $\Sigma | \Delta z_{\text{r}}|$ as the variation of $z_{\text{p}}$ and $z_{\text{r}}$ .
\begin{figure}[hp]
        \centering
        \hfill
        \begin{subfigure}[H]{\linewidth}  
            \centering 
            \includegraphics[width=0.9\linewidth,height=3.6cm,trim=3cm 0.0cm 2cm 2cm]{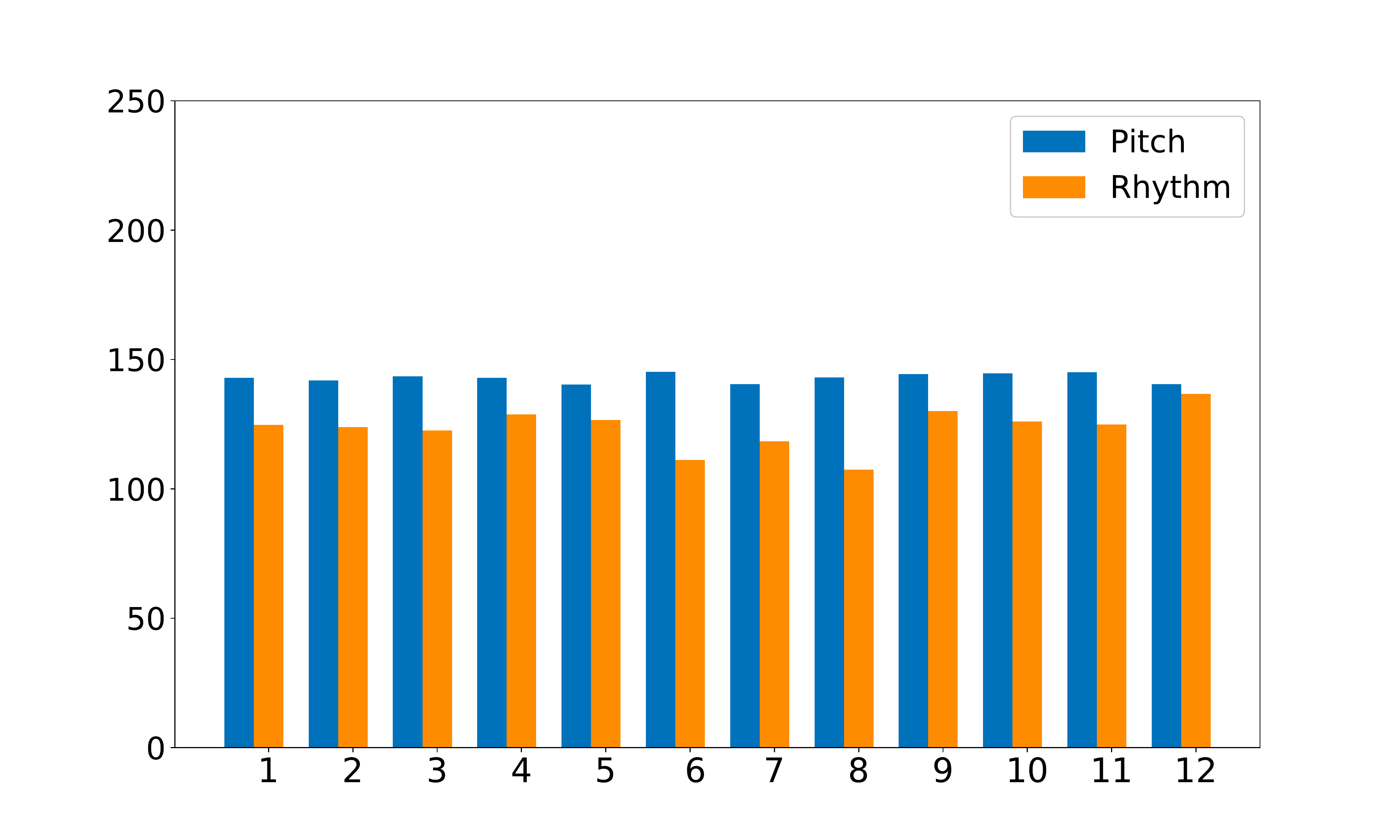}
            \caption{Baseline model (8 bar)}
            \label{fig:base}
        \end{subfigure}
        \begin{subfigure}[H]{\linewidth}  
            \centering
            \includegraphics[width=0.9\linewidth,height=3.6cm,trim=3cm 0.0cm 2cm 0.0cm]{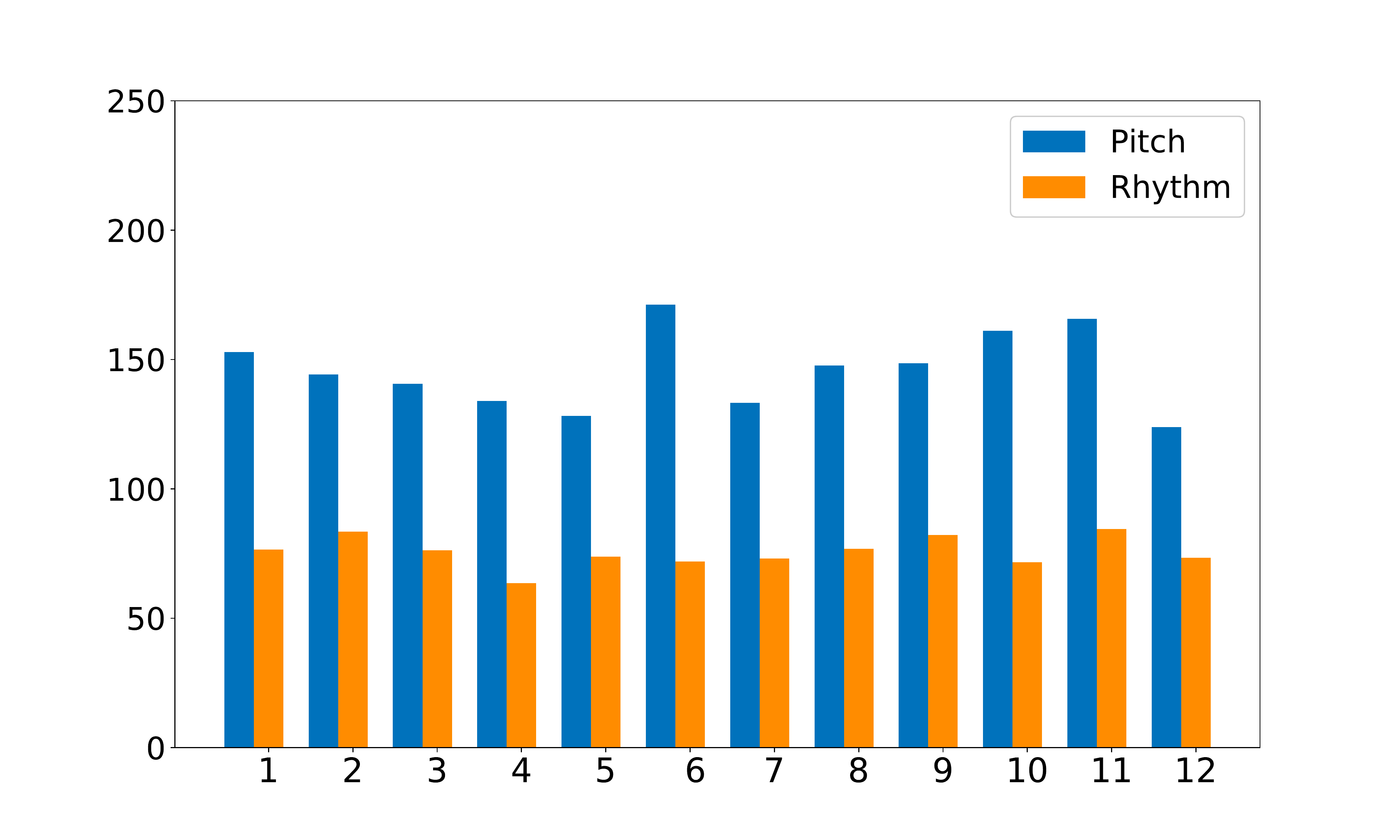}
            \caption{Long-EC$^2$-VAE model (8 bar)}
            \label{fig:compare_bar_kld_0}
        \end{subfigure}
        
        \hfill
        \begin{subfigure}[H]{\linewidth}  
            \centering 
            \includegraphics[width=0.9\linewidth,height=3.6cm,trim=3cm 0.0cm 2cm 0.0cm]{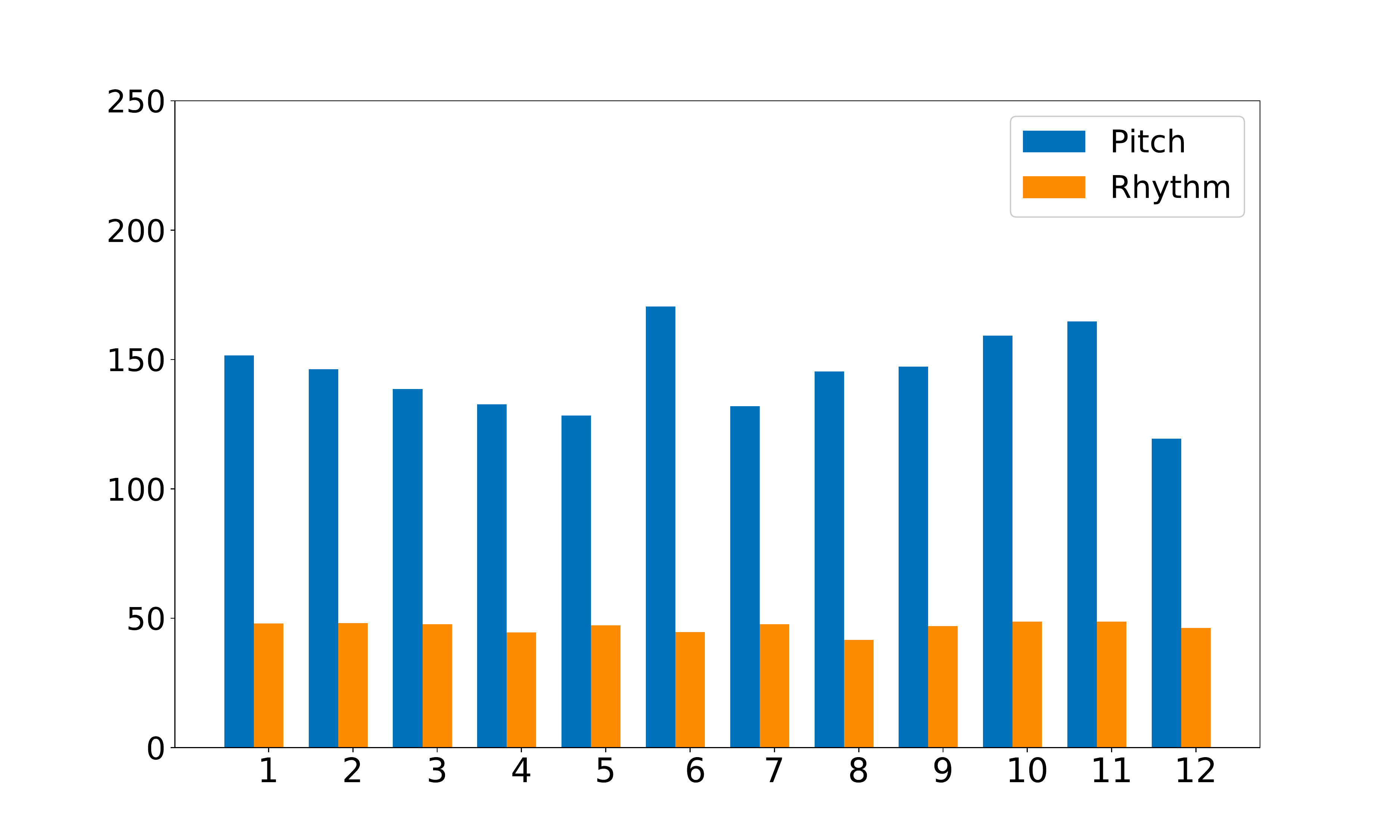}
            \caption{Hirarchical-EC$^2$-VAE model (8 bar)}
            \label{fig:compare_bar_kld_1}
        \end{subfigure}
    \caption{The comparison between $\Sigma|\Delta z_{\rm p}|$ and $\Sigma|\Delta z_{\rm r}|$ after transposition. The numbers show the pitch augmented by 12 semitones in each sub-figure from left to right.}
    \label{fig:disentangle comparison}
    \vspace{-1em}
\end{figure}

 As shown in Figure~\ref{fig:disentangle comparison}, values of $\Sigma|\Delta z_{\text{p}}|$ of the proposed Hierarchical-EC$^{2}$-VAE are relatively high while $\Sigma | \Delta z_{\text{r}}|$ maintains in a significantly low level. This indicates that the pitch and rhythm representations of the proposed Hierarchical-EC$^{2}$-VAE are well-disentangled as the change of notes has a tiny impact on $z_{\text{r}}$.
 Similarly, we can intuitively find in the figure that the disentanglement performance of the proposed Hierarchical-EC$^{2}$-VAE is much better than the baseline and also outperforms the proposed Long-EC$^{2}$-VAE.
\subsection{Music generative examples}
In this section, we show some music generation results by manipulating the disentangled phrase-level pitch and rhythm representations in three different ways: style transfer via swapping the representation, rhythm morphing via interpolating the representation, and theme variation via representation posterior sampling.
\subsubsection{Phrase-level composition style transfer}
 We cross-swap the disentangled pitch and rhythm factors $z_{\rm p}$ and $z_{\rm r}$ of two 8-bar melodies A and B and then obtain generative pieces C and D.
The results are shown in Figure 5, in which we see that both of the two generative pieces perfectly inherit target rhythm patterns. Besides, these generative melodies vary slightly from the source melody and these variations tend to sound creative, i.e. the appearance of embellished notes.

\begin{figure}[htbp]
            \centering 
            \includegraphics[width=\linewidth,clip,height=6.5cm, trim=8cm 2.1cm 7.25cm 0.3cm]{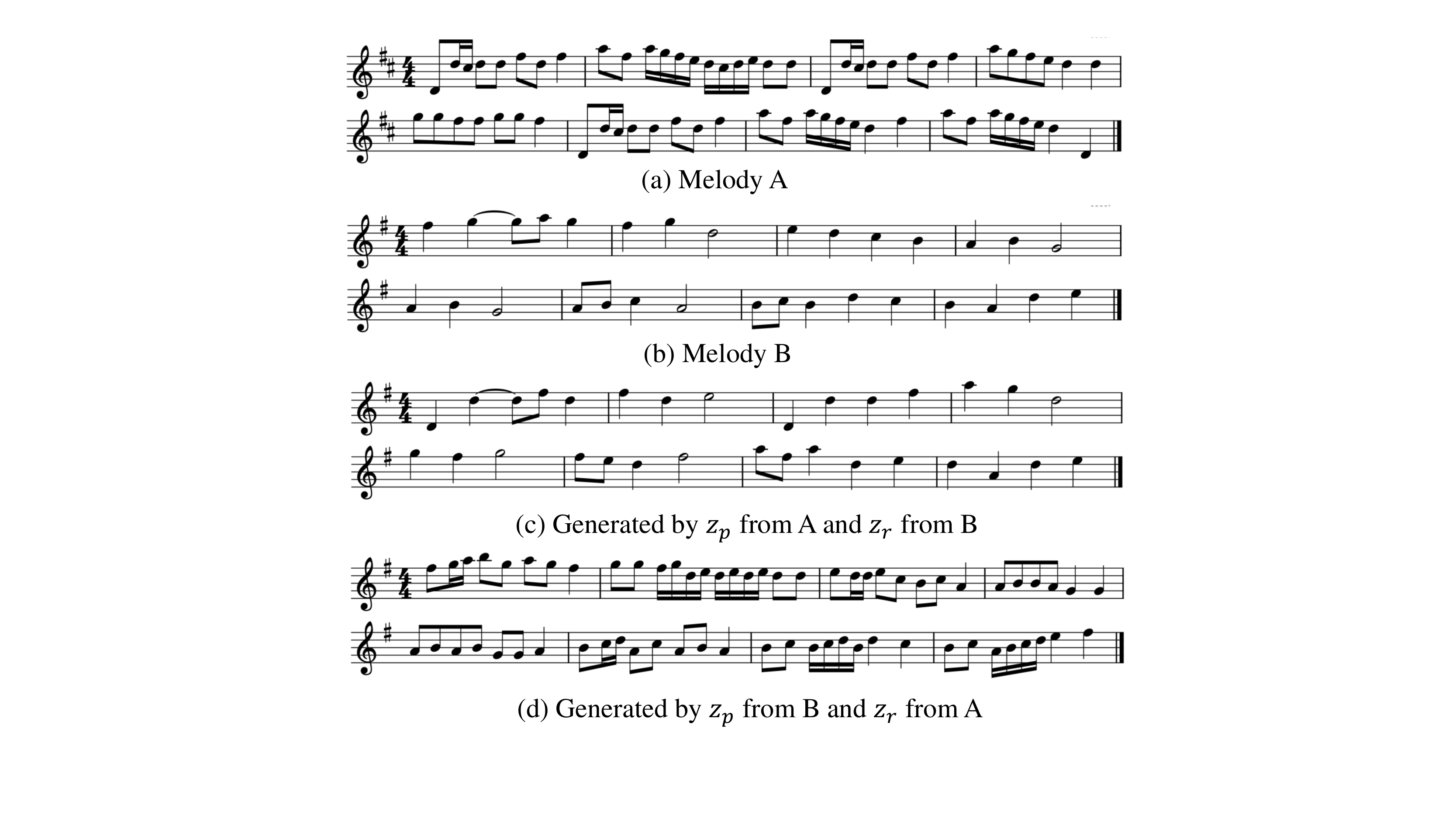}
 \caption{Style transfer examples by hierarchical-EC$^2$-VAE model.}
 \label{fig:wuxian}
\end{figure}
\subsubsection{Latent $z_{\rm r}$ interpolation}
We interpolate rhythm representations $z_{\text{r}}$ of two phrases using SLERP \cite{1994Advanced} while keeping the pitch and chord unchanged. The interpolated latent representations can then be ``re-synthesized'' using Hierarchical-EC$^{2}$-VAE.

\begin{figure}[htb]
            \centering 
            \includegraphics[width=\linewidth,clip, trim=6.5cm 0.5cm 6.3cm 1.5cm]{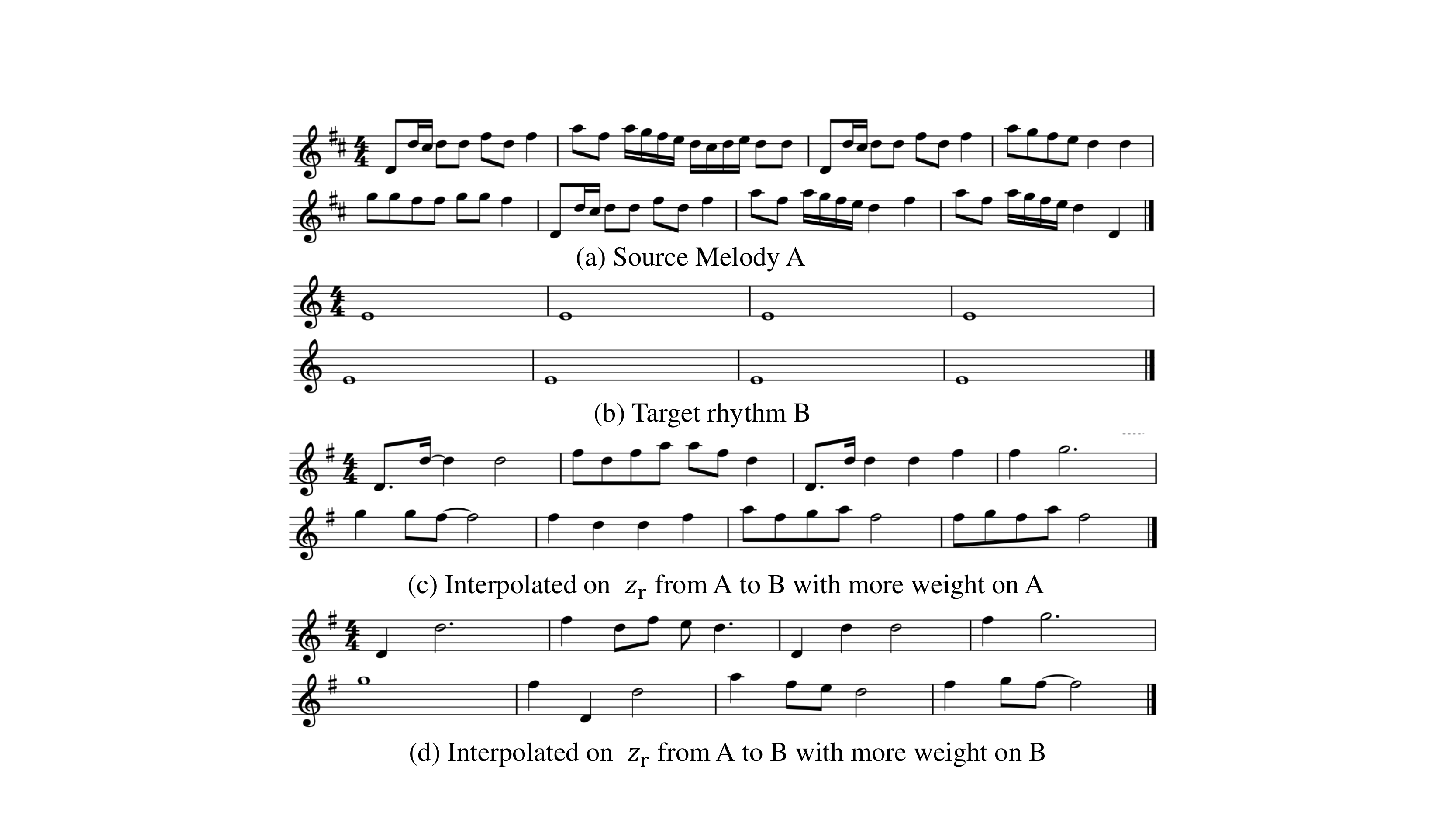}
 \caption{Interpolation examples.}
 \label{fig:interpolation}
\end{figure}

As shown in Figure~\ref{fig:interpolation}, we interpolate $z_{\text{r}}$ of the piece A and B with different SLERP weights. The results exhibit a surprising sense of coherence of pitch and rhythm in generative melodies, even in the transition between consecutive bars.This implies that a longer-term representation is also adept at modeling short-term generation and even contains more global harmonic information than a short-term representation.

\subsubsection{Theme variation}\label{zhuguan}

\begin{figure}[htbp]
            \centering 
            \includegraphics[width=\linewidth,clip, trim=6.5cm 4cm 8.5cm 2cm]{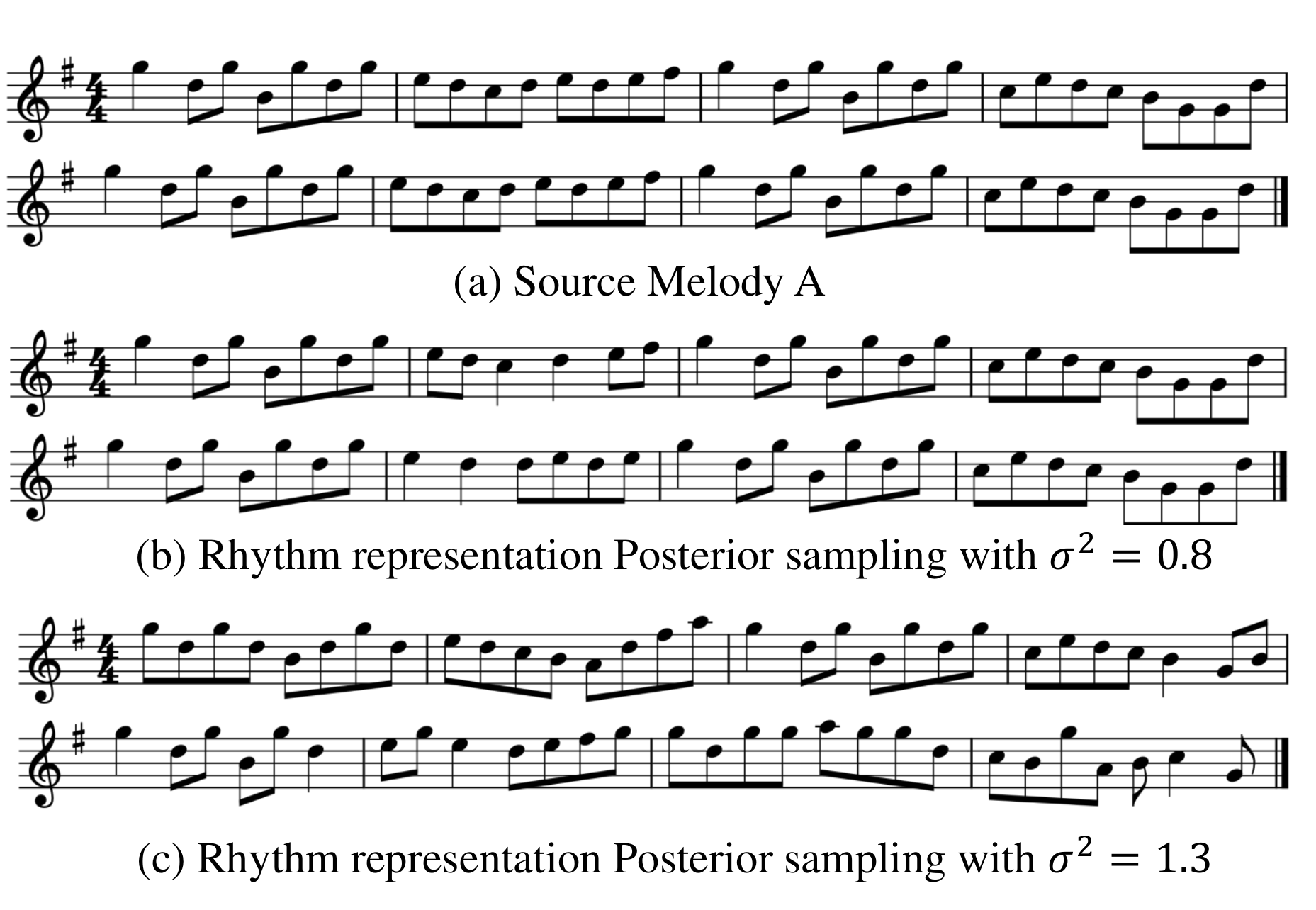}
 \caption{Rhythm representation posterior sampling examples.}
 \vspace{-1em}
 \label{fig:sigma}
\end{figure}

We can also achieve theme variation by adding a Gaussian noise to $z_{\text{r}}$ while keeping $z_{\text{p}}$ unchanged . As a sample shown in Figure~\ref{fig:sigma}, we find that as the variance of the noise grows larger, the pitch and rhythm of the generative melody are still reasonable smooth, implying that the long-term representations contain the coherence of contextual information and can ``control'' the generation process.

\subsection{Subjective Evaluation} 
One may wonder what are the advantages of learning long-term representations since we can always generate the music bar by bar using short-term models and just concatenate the generated samples together. One merit lies in the coherency in controlled music generation. For example, when sampling the long-term rhythm representation, the overall rhythm pattern of a phrase is controlled as an organic whole, while individually sampling the rhythm of different bar may easily lose the rhythm coherency. 
To better illustrate this idea,we conduct a survey on theme variation (as in Section 4.4.3) to compare the performance of the proposed 8-bar Hierarchical-EC$^{2}$-VAE and baseline 2-bar EC$^{2}$-VAE.

\subsubsection{Survey Configuration} 
In our survey, each subject is given 5 groups of pieces. Each group contains three 8 bar pieces: a human-composed piece from Nottingham dataset and 2 theme variations generated by a 2-bar  EC$^{2}$-VAE and Hierarchical-EC$^{2}$-VAE, respectively. In each group, the generated pieces use $z_{\rm p}$ of the human-composed piece and the sampled $z_{\rm r}$. 

Each subject listens to  five randomly arranged groups in turn and is required to rate each melody ranging from 1 (very low) to 5 (very high) according to three aspects: \emph{creativity}, \emph{naturalness} (how human-like the composition is) and overall \emph{musicality}.  
\subsubsection{Results}
\begin{figure}[hbt]
            \centering 
            \includegraphics[width=1\linewidth]{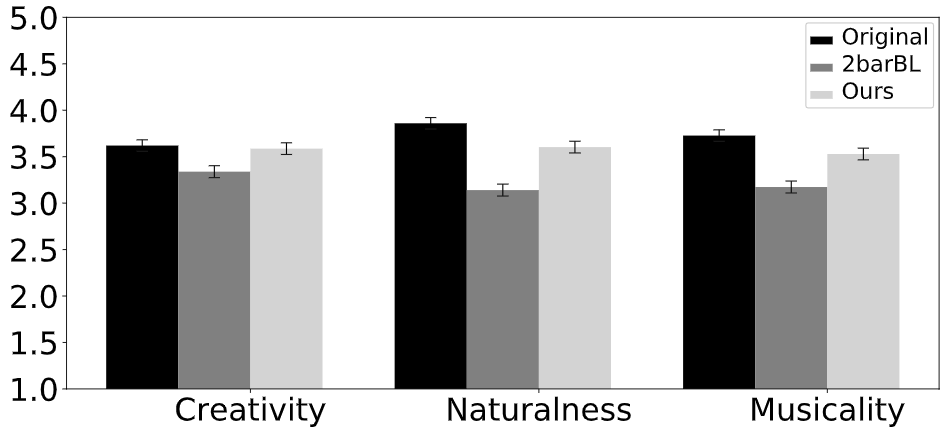}
            \vspace{-1em}
 \caption{The results of the subjective evaluation.}
 \label{zhuguan plot}
\end{figure}

A total of 29 subjects ( 18 females and 11 males) participated in the survey. Experimental results depicted in Figure~\ref{zhuguan plot} demonstrate that people prefer melodies generated by the proposed Hierarchical-EC$^{2}$-VAE to those generated by the 2 bar EC$^{2}$-VAE\cite{DBLP:journals/corr/abs-1906-03626}, implying the effects of a long-term coherence learned by our model. The heights of bars represent means of the ratings and the error bars represent the MSEs computed via within-subject ANOVA \cite{H1999The}. The results show that our model performs significantly better than the 2 bar EC$^2$-VAE in terms of all three dimensions(p < 0.05). Besides, the qualities of melodies generated by the proposed Hierarchical-EC$^{2}$-VAE reach a competitive standard compared to the human-composed pieces, especially in \emph{creativity}.

\vspace{-1em}
\section{Conclusion}
In conclusion, we contribute a pipeline  of algorithms to learn long-term and disentangled music representations. The main novelty lies in the proposed two inductive biases which constrain the long-term representations using contextual information. The first one requires long-term representation to be not too different from the short-term ones which represent a part of the long-term sequence, and we demonstrate contrastive loss is well-suited for such rough constraint. The second inductive bias is that a good long-term representation should be able to reconstruct the corresponding short-term ones, and we use a hierarchical predictive model to realize this constraint. Unlike most hierarchical models, our purpose is not prediction for its own sake, but rather to leverage the prediction power to learn a well-disentangled long-term representation. Experimental results show that our approach is quite successful, capable of disentangling pitch and rhythmic factors for phrase-level (32 beats) melody without increasing the dimensionality of latent representation compared to bar-level models. Moreover, the learned representations enable high-quality phrase-level style transfer via representation swapping and theme variation by representation interpolation and posterior sampling.


\bibliography{main}

\begin{thebibliography}{10}
\providecommand{\url}[1]{#1}
\csname url@samestyle\endcsname
\providecommand{\newblock}{\relax}
\providecommand{\bibinfo}[2]{#2}
\providecommand{\BIBentrySTDinterwordspacing}{\spaceskip=0pt\relax}
\providecommand{\BIBentryALTinterwordstretchfactor}{4}
\providecommand{\BIBentryALTinterwordspacing}{\spaceskip=\fontdimen2\font plus
\BIBentryALTinterwordstretchfactor\fontdimen3\font minus
  \fontdimen4\font\relax}
\providecommand{\BIBforeignlanguage}[2]{{%
\expandafter\ifx\csname l@#1\endcsname\relax
\typeout{** WARNING: IEEEtran.bst: No hyphenation pattern has been}%
\typeout{** loaded for the language `#1'. Using the pattern for}%
\typeout{** the default language instead.}%
\else
\language=\csname l@#1\endcsname
\fi
#2}}
\providecommand{\BIBdecl}{\relax}
\BIBdecl

\bibitem{DBLP:journals/corr/abs-1803-05428}
A.~Roberts, J.~H. Engel, C.~Raffel, C.~Hawthorne, and D.~Eck, ``A hierarchical
  latent vector model for learning long-term structure in music,'' in
  \emph{Proceedings of the 35th International Conference on Machine Learning,
  Stockholm, Sweden}, 2018, pp. 4361--4370.

\bibitem{lstm}
S.~Hochreiter and J.~Schmidhuber, ``Long short-term memory,'' \emph{Neural
  Comput.}, vol.~9, no.~8, pp. 1735--1780, 1997.

\bibitem{attention}
A.~Vaswani, N.~Shazeer, N.~Parmar, J.~Uszkoreit, L.~Jones, A.~N. Gomez,
  L.~Kaiser, and I.~Polosukhin, ``Attention is all you need,'' in
  \emph{Advances in Neural Information Processing Systems 30: Annual Conference
  on Neural Information Processing Systems, Long Beach, CA, USA}, 2017, pp.
  5998--6008.

\bibitem{DBLP:journals/corr/abs-2008-01291}
K.~Chen, G.~Xia, and S.~Dubnov, ``Continuous melody generation via disentangled
  short-term representations and structural conditions,'' in \emph{14th
  International Conference on Semantic Computing, San Diego, CA, USA}, 2020,
  pp. 128--135.

\bibitem{DBLP:journals/corr/abs-2008-07122}
Z.~Wang, D.~Wang, Y.~Zhang, and G.~Xia, ``Learning interpretable representation
  for controllable polyphonic music generation,'' in \emph{Proceedings of the
  21th International Society for Music Information Retrieval Conference,
  Montreal, Canada}, 2020, pp. 662--669.

\bibitem{Inpainting}
A.~Pati, A.~Lerch, and G.~Hadjeres, ``Learning to traverse latent spaces for
  musical score inpainting,'' in \emph{Proceedings of the 20th International
  Society for Music Information Retrieval Conference, Delft, The Netherlands},
  2019, pp. 343--351.

\bibitem{DBLP:journals/corr/abs-1906-03626}
R.~Yang, D.~Wang, Z.~Wang, T.~Chen, J.~Jiang, and G.~Xia, ``Deep music analogy
  via latent representation disentanglement,'' in \emph{Proceedings of the 20th
  International Society for Music Information Retrieval Conference, Delft, The
  Netherlands}, 2019, pp. 596--603.

\bibitem{poptransformer}
Y.~Huang and Y.~Yang, ``Pop music transformer: Beat-based modeling and
  generation of expressive pop piano compositions,'' in \emph{{MM} '20: The
  28th {ACM} International Conference on Multimedia, Virtual Event / Seattle,
  WA, USA}, 2020, pp. 1180--1188.

\bibitem{midivae}
G.~Brunner, A.~Konrad, Y.~Wang, and R.~Wattenhofer, ``{MIDI-VAE:} modeling
  dynamics and instrumentation of music with applications to style transfer,''
  in \emph{Proceedings of the 19th International Society for Music Information
  Retrieval Conference, Paris, France}, 2018, pp. 747--754.

\bibitem{DBLP:conf/ismir/Akama19}
T.~Akama, ``Controlling symbolic music generation based on concept learning
  from domain knowledge,'' in \emph{Proceedings of the 20th International
  Society for Music Information Retrieval Conference, Delft, The Netherlands},
  2019, pp. 816--823.

\bibitem{DBLP:conf/ismir/LuoAH19}
Y.~Luo, K.~Agres, and D.~Herremans, ``Learning disentangled representations of
  timbre and pitch for musical instrument sounds using gaussian mixture
  variational autoencoders,'' in \emph{Proceedings of the 20th International
  Society for Music Information Retrieval Conference, Delft, The Netherlands},
  2019, pp. 746--753.

\bibitem{9054554}
J.~Jiang, G.~Xia, D.~B. Carlton, C.~N. Anderson, and R.~H. Miyakawa,
  ``Transformer {VAE:} {A} hierarchical model for structure-aware and
  interpretable music representation learning,'' in \emph{International
  Conference on Acoustics, Speech and Signal Processing, Barcelona, Spain},
  2020, pp. 516--520.

\bibitem{DBLP:conf/icml/ChenK0H20}
T.~Chen, S.~Kornblith, M.~Norouzi, and G.~E. Hinton, ``A simple framework for
  contrastive learning of visual representations,'' in \emph{Proceedings of the
  37th International Conference on Machine Learning}, 2020, pp. 1597--1607.

\bibitem{DBLP:conf/cvpr/He0WXG20}
K.~He, H.~Fan, Y.~Wu, S.~Xie, and R.~B. Girshick, ``Momentum contrast for
  unsupervised visual representation learning,'' in \emph{Conference on
  Computer Vision and Pattern Recognition, Seattle, WA, USA}, 2020, pp.
  9726--9735.

\bibitem{DBLP:journals/corr/abs-2003-04297}
X.~Chen, H.~Fan, R.~B. Girshick, and K.~He, ``Improved baselines with momentum
  contrastive learning,'' \emph{CoRR}, vol. abs/2003.04297, 2020.

\bibitem{ncebased}
M.~Gutmann and A.~Hyv{\"{a}}rinen, ``Noise-contrastive estimation: {A} new
  estimation principle for unnormalized statistical models,'' in
  \emph{Proceedings of the Thirteenth International Conference on Artificial
  Intelligence and Statistics}, ser. {JMLR} Proceedings, vol.~9, 2010, pp.
  297--304.

\bibitem{WORDNCE}
A.~Mnih and K.~Kavukcuoglu, ``Learning word embeddings efficiently with
  noise-contrastive estimation,'' in \emph{Advances in Neural Information
  Processing Systems 26: 27th Annual Conference on Neural Information
  Processing Systems, Lake Tahoe, Nevada, United States}, 2013, pp. 2265--2273.

\bibitem{DBLP:journals/corr/abs-1807-03748}
A.~van~den Oord, Y.~Li, and O.~Vinyals, ``Representation learning with
  contrastive predictive coding,'' \emph{CoRR}, vol. abs/1807.03748, 2018.

\bibitem{DBLP:conf/iclr/HjelmFLGBTB19}
R.~D. Hjelm, A.~Fedorov, S.~Lavoie{-}Marchildon, K.~Grewal, P.~Bachman,
  A.~Trischler, and Y.~Bengio, ``Learning deep representations by mutual
  information estimation and maximization,'' in \emph{7th International
  Conference on Learning Representations,New Orleans, LA}, 2019.

\bibitem{1996A}
F.~Lerdahl and R.~Jackendoff, ``A generative theory of tonal music,''
  \emph{Journal of Aesthetics and Art Criticism}, vol.~9, no.~1, pp. 72--73,
  1996.

\bibitem{1983Introduction}
W.~Rothstein, O.~Jonas, and J.~Rothgeb, ``Introduction to the theory of
  heinrich schenker: The nature of the musical work of art,'' \emph{Journal of
  Music Theory}, vol.~27, no.~2, 1983.

\bibitem{2006Implementing}
M.~Hamanaka, K.~Hirata, and S.~Tojo, ``Implementing "a generative theory of
  tonal music",'' \emph{Journal of New Music Research}, vol.~35, no.~4, pp.
  249--277, 2006.

\bibitem{Schenkerian}
A.~Marsden, ``Schenkerian analysis by computer: A proof of concept,''
  \emph{Journal of New Music Research}, vol.~39, no.~3, pp. 269--289, 2010.

\bibitem{MusicFaderNets}
H.~H. Tan and D.~Herremans, ``Music fadernets: Controllable music generation
  based on high-level features via low-level feature modelling,'' in
  \emph{Proceedings of the 21th International Society for Music Information
  Retrieval Conference, Montreal, Canada}, 2020, pp. 109--116.

\bibitem{VQmusic}
G.~Hadjeres and L.~Crestel, ``Vector quantized contrastive predictive coding
  for template-based music generation,'' \emph{CoRR}, vol. abs/2004.10120,
  2020.

\bibitem{dhariwal2020jukebox}
P.~Dhariwal, H.~Jun, C.~Payne, J.~W. Kim, A.~Radford, and I.~Sutskever,
  ``Jukebox: {A} generative model for music,'' \emph{CoRR}, vol.
  abs/2005.00341, 2020.

\bibitem{Nottingham}
E.~Foxley, ``Nottingham database,'' 2011.

\bibitem{DBLP:journals/corr/abs-2008-07142}
Z.~Wang, K.~Chen, J.~Jiang, Y.~Zhang, M.~Xu, S.~Dai, and G.~Xia, ``{POP909:}
  {A} pop-song dataset for music arrangement generation,'' in \emph{Proceedings
  of the 21th International Society for Music Information Retrieval Conference,
  Montreal, Canada}, 2020, pp. 38--45.

\bibitem{DBLP:journals/corr/KingmaB14}
D.~P. Kingma and J.~Ba, ``Adam: {A} method for stochastic optimization,'' in
  \emph{3rd International Conference on Learning Representations}, Y.~Bengio
  and Y.~LeCun, Eds., 2015.

\bibitem{DBLP:conf/icml/KimM18}
H.~Kim and A.~Mnih, ``Disentangling by factorising,'' in \emph{Proceedings of
  the 35th International Conference on Machine Learning}, 2018, pp. 2654--2663.

\bibitem{1994Advanced}
S.~Hollasch, ``Advanced animation and rendering techniques: By alan watt and
  mark watt, acm press,'' \emph{Computers \& Graphics}, vol.~18, no.~2, p. 249,
  1994.

\bibitem{H1999The}
H.~Scheffé, ``The analysis of variance,'' in \emph{Architectural Institute of
  Japan}, 1999.

\end{thebibliography}

\end{document}